\documentclass[a4paper]{spie}  
\usepackage[dvips]{graphicx}
\usepackage{amssymb, amsmath, subfigure}
\providecommand{\norm}[1]{\lVert#1\rVert}

\title{Signal processing and statistical methods in analysis of
  text and DNA} 

\author{Matthew J. Berryman\supit{a}, Andrew Allison\supit{a}, Pedro Carpena\supit{b}, 
and Derek Abbott\supit{a}
\skiplinehalf
\supit{a}Center for Biomedical Engineering and\\ 
School of Electrical and Electronic Engineering,\\
The University of Adelaide, SA  5005, Australia\\
\supit{b}Departmenta de Fisica Aplicada II,\\
Universidad de M\'{a}laga, E-29071, Spain
}

\authorinfo{Further author information: (Send correspondence to Derek Abbott)\\
  Derek Abbott: E-mail: dabbbott@eleceng.adelaide.edu.au, Telephone: +61 8 8303 5748}

\begin{document} 
\maketitle 
  
\begin{abstract}
  A number of signal processing and statistical methods can be used in
  analyzing either pieces of text or DNA sequences. These techniques can be used
  in a number of ways, such as determining authorship of documents, finding genes
  in DNA, and determining phylogenetic and linguistic trees. Signal processing
  methods such as spectrograms provide useful new tools in the area of genomic
  information science. In particular, fractal analysis of DNA ``signals'' has
  provided a new way of classifying organisms.
\end{abstract}


\section{INTRODUCTION}
The Human Genome Project~\cite{HSapiens} together with a number of other projects has 
produced the DNA sequences for a large number of organisms, from humans and mice, to zebrafish,
yeast, and over eighty bacteria. There has been a great deal of work done in applying 
signal processing and statistical methods to DNA recently, and our work has been looking at
the application of these methods to not only DNA but in the field of text analysis as well.

A number of interesting statistical techniques have been explored in recent times in the area
of text analysis. These attempt to answer questions regarding what the text
is about (by extracting relevant keywords)~\cite{Keywords} or relationships between texts
or languages~\cite{Zipping}. In this paper we show significant results of analysis performed
on ancient Greek texts, and give a detailed explanation of an F-statistic method for
keyword extraction we have been developing.

In the field of DNA analysis, techniques such as the discrete Fourier
transform~\cite{GSP} and multifractal analysis~\cite{multifractal1} have been explored. In
this paper we present exciting new applications of those methods to the areas of
sequence analysis~\cite{SequenceToFunction} and phylogenetic trees~\cite{protein_trees} (those
showing the relationships between organisms) respectively.
\section{TEXT ANALYSIS}
\subsection{Introduction to text analysis}
In this section we explore new methods from two main areas of text analysis, namely stylography
and keyword extraction. Stylography or more general style analysis using statistical methods
can highlight differences or similarities between authors, and between languages. Although
the idea of using statistical properties of texts in extracting keywords and
categorizing texts is not new~\cite{Bookstein1,Bookstein2}, here
we use a new method of analyzing the spacing between words for both determining style 
similarities and finding keywords. We begin by reviewing the seminal work of
Ortu\~{n}o {\it et al.}~\cite{Keywords}, and then introduce a new keyword extraction
technique based on an F-statistic method.
\subsection{Stylography}
Ortu\~{n}o {\it et al.}~\cite{Keywords} suggest using standard deviation of the inter-word 
spacing to characterize word distributions and extract keywords, as opposed to using a 
frequency count of each word. By inter-word spacing, we mean the number of words in between 
successive occurrences of a keyword (non-inclusive), for example if the keyword is ``the'', 
then the spacing in ``The cat sat on the mat'' is three, and the spacing in 
``The cat is the best cat.'' is two as there are two and three words respectively 
between the two occurrences of the word ``the''.
Initial results of plotting the standard deviation of word
spacing for (almost) all the words in a given text against the plot of those from other texts 
by the same author revealed an unexpected result; namely that works by the same author have a 
similar distribution of inter-word spacings.
A striking result is obtained when we plot the gospels of {\it Matthew} and {\it Luke}, and 
the book of {\it Acts} from the {\it Koine} Greek New Testament. This suggests a statistical 
approach to stylography could be taken, and we demonstrate the use of this for both texts by 
known authors, and for texts where the historical authorship is unclear.
The following method can be used to give an indication of the similarity in style between two 
texts.
We conjecture that this can give a valuable insight into the authorship of texts.

Given a set of word spacings $\{x_{1},\ldots,x_{n}\}$ for a given word, we compute the scaled
standard deviation of word spacings
\begin{equation}
\hat{\sigma} = \frac{1}{\bar{x}}
\sqrt{\displaystyle \sum_{i=1}^{n} \frac{\left(x_{i}-\bar{x}\right)^{2}}{n-1}}.
\label{stddev}
\end{equation}

\begin{figure}[htbp]
\centering{\resizebox{10cm}{!}{\includegraphics[angle=270]{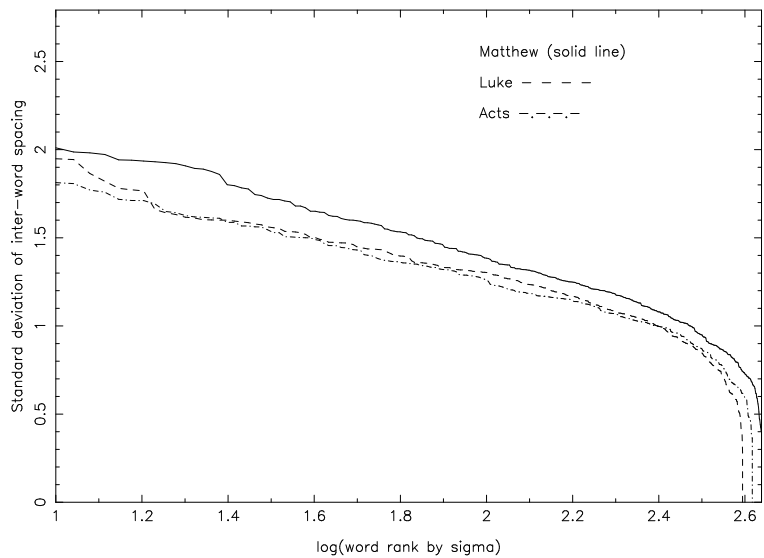}}}
\caption{The scaled standard deviation of the inter-word spacing (y-axis) for each word is 
ranked in descending order on a logarithmic scale (x-axis). Using the original {\it Koine}
Greek text, a remarkably close match is obtained between the gospel of {\it Luke} and the book
of {\it Acts} in the New Testament, which were written by the same author. For reference,
a curve of a different author is shown (the book of {\it Matthew}) illustrating a distinct 
difference (this is the upper curve). Although the match between {\it Luke} and {\it Acts} 
deviates for a log rank $<1.2$, this represents less than four per cent of the total curve 
(due to the base-ten logarithmic scale). Note that uncommon words occurring less than 5 times 
in each text are not included in the ranking, as their scaled standard deviation values are not 
significant.}
\label{gagraph}
\end{figure}

\begin{figure}[htbp]
\centering{\resizebox{10cm}{!}{\includegraphics[angle=270]{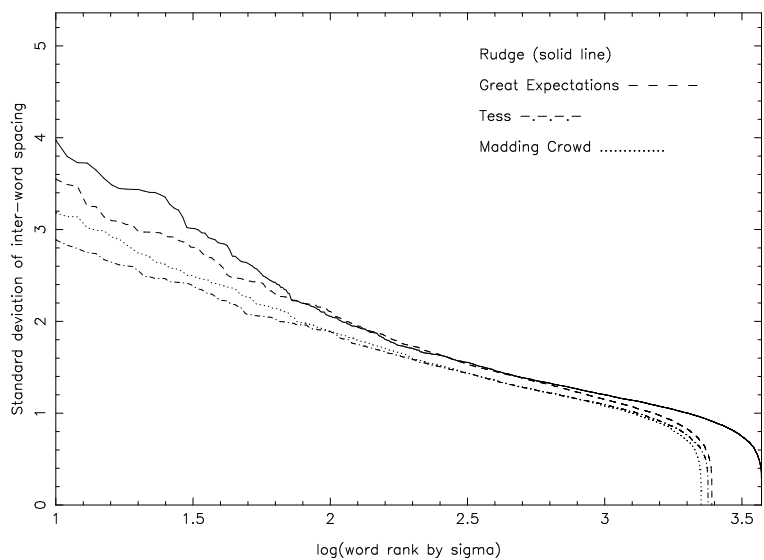}}}
\caption{Standard deviation vs.~log(rank) for books by Charles Dickens and Thomas Hardy. For much
of the length of the plots the graphs for the Dickens texts are nearly coincident, and likewise
for the pair of Hardy texts. Although the plots are all quite different for the region 1 to 1.7, 
note that this apparently large region is quite small due the logarithmic scale on the x-axis. 
The extra length on the plot for {\it Barnarby Rudge} is simply due to the larger number of 
different words in this text when compared with the other three texts.} 
\label{dhgraph}
\end{figure}

We repeat this calculation for all the words in a text, giving us a set of standard deviations 
$\{\hat{\sigma}_{1},\ldots,\hat{\sigma}_{m}\}$. In order to generate the graphs we then rank 
these $\hat{\sigma}_{j}$ in order and plot standard deviation 
vs.~$\log_{10}\left({\rm rank}\right)$. 
We omit those words occurring five or fewer times as being statistically insignificant.
The technique of using inter-word spacing in this way and plotting the ranked standard 
deviations on a logarithmic scale was suggested by Carpena 
{\it et al.}~\cite{Carpena}, however here 
we show some striking similarities
between graphs of standard deviations of words in texts by the same author.
Figure~\ref{gagraph} shows the result of applying this to the gospels of {\it Matthew} 
and {\it Luke}, and the book of {\it Acts}.  
Note that we have used the {\it Koine} Greek sources for the New Testament~\cite{Koine}
to eliminate any changes in style due to translation.
Figure~\ref{dhgraph} shows the similarity between works by Charles Dickens ({\it Great
Expectations} and {\it Barnaby Rudge}) and works by Thomas Hardy 
({\it Tess of the d'Urbervilles} and 
{\it Far From the Madding Crowd}). 
This illustrates that the technique is not restricted to only Greek texts.
\subsection{Keyword extraction}
Here we detail the method used by Ortu\~{n}o {\it et al.}~\cite{Keywords} and introduce a 
new method for extracting keywords.

As per the standard deviation graphs, 
we determine the set of standard deviations of word spacings for all
the words in a text, $\{\hat{\sigma}_{1},\ldots,\hat{\sigma}_{m}\}$. Again, we rank the 
words from highest standard deviation from highest to lowest, but keeping all the words. 
We thus obtain a list of words 
ranked from high relevance to low relevance.

Another method we have examined for keyword extraction is using the F-statistic on the word 
spacings, assuming a geometric distribution.
The F-statistic detects word-spacing with excess variance (relative to a maximal-entropy or 
``geometric'' distribution).
The F-statistic behaves asymptotically like a Gaussian random variable (when the
number of inter-word spacing samples is large)  with mean of 0 and variance of 1
so the statistical tests for relevant keywords are very easy.
Given the set of word spacings $\{x_{1},\ldots,x_{n}\}$ we use the F-statistic
\begin{equation}
\frac{1}{2}\ln(n)\left(\frac{s^{2}}{\bar{x}(1+\bar{x})}-1\right),
\label{Fstat}
\end{equation}
where $s$ is the normal sample standard deviation. Note the similarity between the F-statistic 
and the square of the scaled standard deviation. 
Assuming the null hypothesis then we get a maximum likelihood estimate of the parameter 
$a$ in the geometric pdf $p(x) = (1-a)(a^{x})$, and can hence estimate the variance of the 
distribution and we compare this with the standard unbiased estimator for variance. 
The $\log(n)$ term is for scaling (to deal with the accuracy of the F-statistic for 
different sample sizes). Other terms are corrections as detailed in Abramowitz and 
Stegun~\cite{Abramowitz}.

Table~\ref{rankings} shows words from Sir Arthur Conan Doyle's 
{\it The Hound of the Baskervilles}.
\begin{table}[htb]
\caption{Rankings according to frequency, standard deviation and F-statistic for words in the
{\it The Hound of the Baskervilles}.}
\centering\footnotesize
\begin{tabular}{|c|c|c|c|c|c|c|}\hline
{\bf Rank} & {\bf Freq. word} & {\bf Freq. val.} & {\bf $\hat{\sigma}$ word} &
{\bf $\hat{\sigma}$ val.} & {\bf F-stat. word} & {\bf F-stat. val.} \\\hline
{\bf 1} & the & 3327 & her & 3.23 & her & 24.02\\\hline
{\bf 2} & and & 1628 & hotel & 3.11 & she & 17.83\\\hline
{\bf 3} & of & 1592 & she & 2.88 & we & 15.08\\\hline
{\bf 4} & i & 1465 & mortimer & 2.75 & mortimer & 14.54\\\hline
{\bf 5} & to & 1448 & we & 2.44 & hotel & 13.96\\\hline
{\bf 6} & a & 1306 & hound & 2.41 & you & 10.82\\\hline
{\bf 7} & that & 1133 & body & 2.38 & hound & 10.18\\\hline
{\bf 8} & it & 979 & boot & 2.31 & charles & 6.92\\\hline
{\bf 9} & he & 914 & hugo & 2.28 & body & 6.77\\\hline
{\bf 10} & in & 913 & alley & 2.19 & barrymore & 6.61\\\hline
{\bf 11} & you & 826 & heir & 2.15 & boot & 6.41\\\hline
{\bf 12} & was & 803 & cab & 2.13 & i & 6.08\\\hline
{\bf 13} & his & 689 & high & 2.13 & hugo & 5.84\\\hline
{\bf 14} & is & 622 & miss & 2.11 & your & 5.62\\\hline
{\bf 15} & have & 541 & you & 2.07 & cab & 5.38\\\hline
\end{tabular}

\label{rankings}
\end{table}
As can be seen in Table~\ref{rankings}, the standard deviation and F-statistic methods are 
much better than a simple frequency count, which tends only to pick out the conjunctive words.

Qualitative tests need to be carried out to establish whether the standard deviation or 
F-statistic method performs best at extracting relevant keywords. We are currently working
on a web search engine which ranks pages by the score of the keywords the user is searching
for (using any of the methods described above or a combination of them).

\section{DNA ANALYSIS}
\subsection{Introduction to DNA analysis}
DNA is like text, but instead of coding for human thoughts it codes for the building
blocks of all living things. Here we present several new techniques for analyzing
regions of DNA and classifying organisms based on their whole genome.
For the purposes of our analysis, we simply treat DNA as a long string with letters
from an alphabet $A={a,t,c,g}$. We then map that alphabet into numerical sequences, 
and then use standard signal processing and statistical methods to analyze those sequences.

The first method shows how color spectrograms give a visual feel for various properties
of the DNA sequences. The second method then explores how multifractal methods
and spacing methods like we have used for text analysis above can be used in
classifying bacteria.

\subsection{Analysis of coding regions in DNA using color spectrograms}
Color spectrograms are a useful tool in visualizing aspects of signals occurring in time,
for example one can see the noise present in the audio recording of the moon landing
and then design an efficient filter to remove it, using only the visual information 
provided in the spectrogram to determine the noise.

Distinguishing coding from non-coding regions is an important problem in genetics. Others
such as Bernaola-Galv\'{a}n {\it et al.} have explored entropy-based methods for
separating the coding from the non-coding regions~\cite{Borders}. Can color spectrograms
give a visual guide to these regions? Here we show some results which suggest this is possible.

Anastassiou~\cite{GSP} has explored the use of color spectrograms in analyzing DNA
sequences. As detailed further in Anastassiou, for a sequence of bases numbered
$1,\ldots,n,\ldots,N$, we can define the following sequences:
\begin{equation}
\begin{array}{c}
x_{r}[n]=\frac{\sqrt{2}}{3}(2u_{T}[n]-u_{C}[n]-u_{G}[n]),\\\\
x_{g}[n]=\frac{\sqrt{6}}{3}(u_{C}[n]-u_{G}[n]),\\\\
x_{b}[n]=\frac{1}{3}(3u_{A}[n]-u_{T}[n]-u_{C}[n]-u_{G}[n]),
\end{array}
\label{colortime}
\end{equation}
where $u_{X}[n]=1$ if the base at position $n$ is $X$, zero otherwise.
The sequences $x_{r},x_{g},x_{b}$ are used in generating red, green, and blue color components
of pixels (the squares) in the spectrograms. The mapping of a base at position $n$, from
the set $\{a,t,c,g\}$ onto the sequences $x_{r},x_{g},x_{b}$ is done so as to maximize the
differences between the sequences at that position $n$, which results in more vivid 
colorings of the spectrogram.
To color the spectrogram we compute the discrete Fourier transform (DFT)
\begin{equation}
X_{c}[k]=\displaystyle \sum_{n=0}^{59}x_{c}[l]e^{-2\pi j n k /60},
\label{DFT}
\end{equation}
for $k=1,\ldots,30$ and where $c$ represents one of the color sequences. Note we do not scale 
the DFT expression in Eq.~\ref{DFT} with a $1/N$ term as usual, as we have to scale the resulting
$\lvert X_{c}[k] \rvert$ into the color component range $\{0,\ldots,255\}$. 
We use a DFT block size of 60 as opposed to a power of two (which would enable us to use the FFT
algorithm to improve the computational complexity) since 60 has a large number of integer 
dividers, some of which correspond to common repeat lengths in DNA, for example the frequency 
$k=20$ corresponds to the codon length $3=60/20$. Repeats of length two and six are also 
common in sections of DNA, these have frequencies of $k=60/2=30$ and $k=60/2=30$. These 
repeat lengths thus give rise to center-cell frequencies, so there is no sidelobe leakage
for these repeat lengths.

\begin{figure}[htbp]
  \centering
  \subfigure[Color spectrogram of a gene in {\it S.~aureus}]{
    \label{spectrograms:staph}
    \includegraphics[width=6cm]{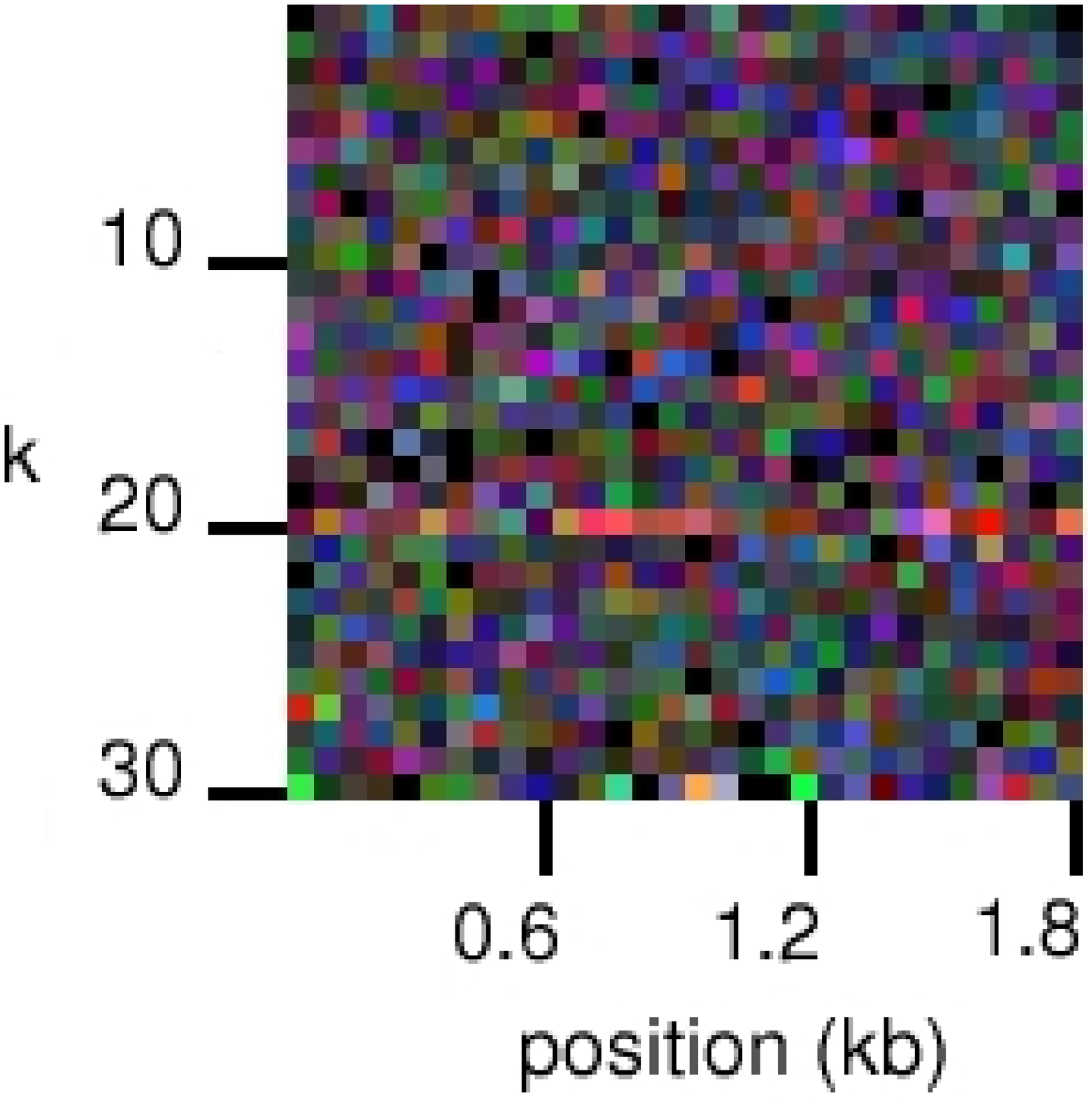}
  }
  \subfigure[Color spectrogram of a gene in {\it H.~sapiens}]{
    \label{spectrograms:us}
    \includegraphics[width=6cm]{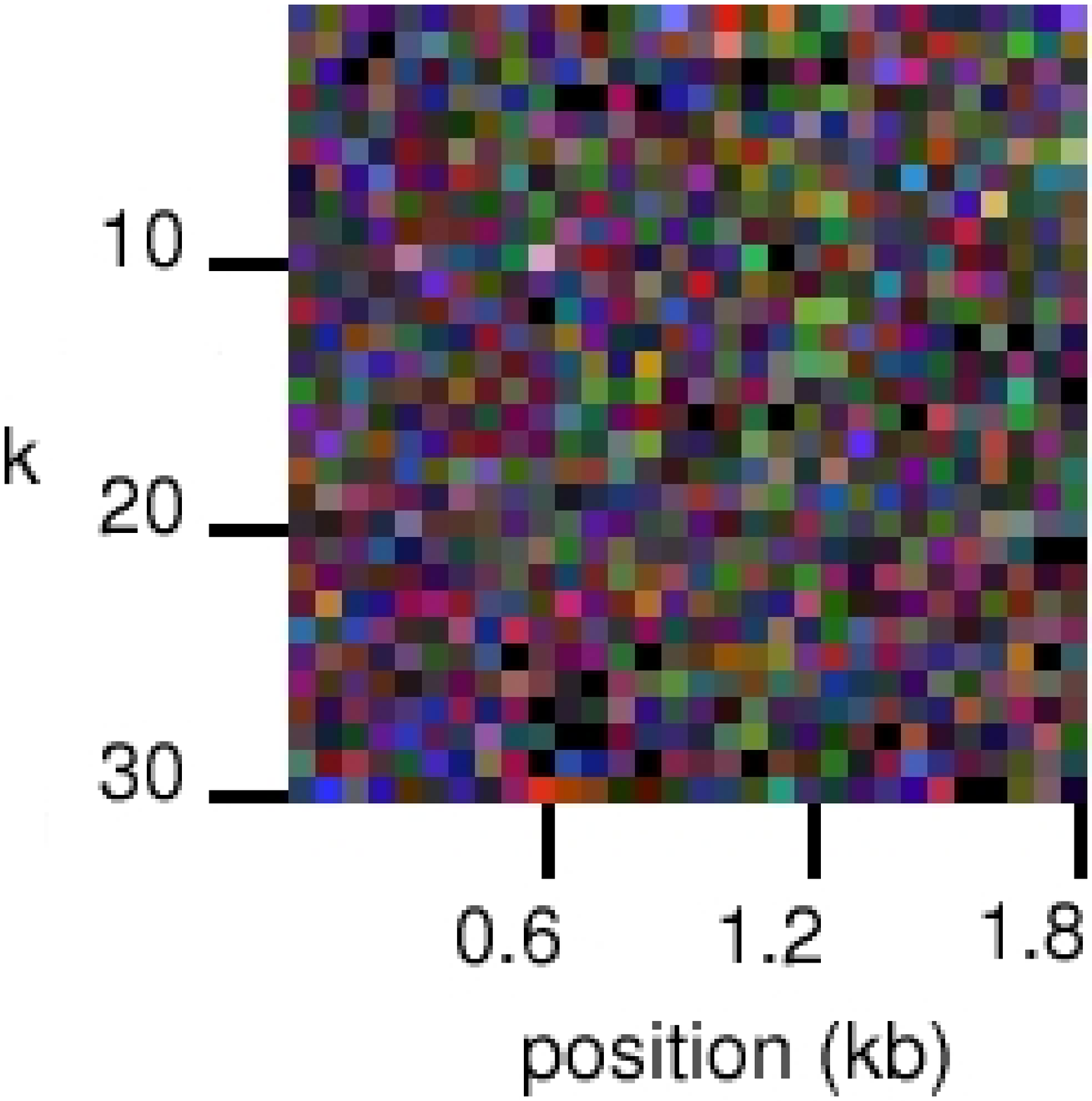}
  }
  \caption{Figure~\ref{spectrograms:staph} shows the color spectrogram for a 1.8 kb 
    region in the {\it gyrA} gene of {\it S.~aureus Mu50}. Note the band at a
    frequency $k=20$, indicating a section with codons occurring repeatedly in a long 
    sequence. This differs to Figure~\ref{spectrograms:us}, which shows a 1.8 kb region
    in the {\it Q9BZA9} gene of {\it H.~sapiens}, indicating less repetition of codons, 
    perhaps because of the presence of introns.}
  \label{spectrograms}
\end{figure}

To illustrate the usefulness of this technique in identifying regions of DNA, 
we show in Figure~\ref{spectrograms} the color spectrograms of DNA sequences in 
{\it Staphylococcus aureus Mu50}~\cite{SAureus} and {\it Homo sapiens}~\cite{HSapiens}.
Figure~\ref{spectrograms} suggests there are differences in the spectrograms between coding
regions of DNA and regions with both coding and non-coding regions. The lack of fine-grained
resolution of the spectrograms is problematic, and prevents easy visualization of the borders
between coding and non-coding regions.
\subsection{Multifractal analysis}
A method useful in comparing different organisms is to use a
multifractal method.
We use the fractal method as detailed by Yu {\it et al.}~\cite{Multifractal2}
namely to each possible substring $s=s_{1} \ldots s_{k}$, $s_{i} \in A$ 
of DNA of length $K$, 
we assign a unique set $[x_{l},x_{r})$ given by
\begin{equation}
x_{l}(s)=\displaystyle \sum_{i=1}^{K} \frac{x_{i}}{4^{i}},
\label{xl}
\end{equation}
where
\begin{equation}
x_{i}=
\begin{cases}
0, & s_{i}=a,\\
1, & s_{i}=c,\\
2, & s_{i}=g,\\
3, & s_{i}=t,
\end{cases}
\label{xi}
\end{equation}
and
\begin{equation}
x_{r}(s)=x_{l}(s)+\frac{1}{4^{K}}.
\label{xr}
\end{equation}
Then
\begin{equation}
F(s)=\frac{N(s)}{L-K+1},
\label{F}
\end{equation}
where $N(s)$ is the number of occurrences of the substring $s$ in the string of length $L$ 
of the whole genome. The fractal measure is then
\begin{equation}
\mu_{K}(dx)=Y_{K}(x)dx,
\label{measure}
\end{equation}
where
\begin{equation}
Y_{K}(x)=4^{K}F_{K}(s), x\in[x_{l}(s),x_{r}(s)).
\label{Y}
\end{equation}
The partition sum is
\begin{equation}
Z_{\epsilon}(q)=
\begin{cases}
\displaystyle \sum_{\mu(B) \neq 0}[\mu(B)]^{q}, & q \neq 1,\\
\\
\displaystyle \sum_{\mu(B) \neq 0}\mu(B)\ln \mu(B), & {\rm otherwise}.
\end{cases}
\label{Z}
\end{equation}
Here we run over all non-empty boxes $B=[n\epsilon,(n+1)\epsilon)$
where $\epsilon$ is $\epsilon=4^{-K}$ and $n=1,\ldots,4^{K}-1$.
Since $\mu(B) \in {\mathbf R}$ and addition is commutative in the reals, the ordering of the 
$\mu(B)$ given by Eq.~\ref{xi} is unimportant in calculating Eq.~\ref{Z}. It is therefore
unimportant in calculating the R\'{e}nyi dimension $D_{q}$ for $q \in {\mathbf R}$, 
given by
\begin{equation}
D_{q}=
\begin{cases}
\displaystyle \lim_{\epsilon \to 0}\frac{\ln Z_{\epsilon}(q)}{(q-1)\ln \epsilon}, & q \neq 1,\\
\\
\displaystyle \lim_{\epsilon \to 0}\frac{Z_{\epsilon}(q)}{\ln \epsilon}, & q = 1
\end{cases}
\label{Renyi}
\end{equation}
Note that although the method used by Yu {\it et al.} doesn't show long-range
correlations in the DNA sequence, we are considering the information content in 
the sequence and not the correlations. If you consider the case where $q=1$, then the 
R\'{e}nyi dimension $D_{1}$ is the same as the Shannon entropy~\cite{ShannonPapers}.
As differences in and similarities in G+C content can indicate relationships between
organisms~\cite{Xanthomonas}, here we are using the R\'{e}nyi dimensions to 
determine if this is reflected in a useful way in an uneven
distribution of the segments -- the ordering is unimportant here, since we are only
comparing the unevenness of the distribution, and not properties relating to the ordering.

\begin{figure}[htbp]
\centering{\resizebox{10cm}{!}{\includegraphics[angle=270]{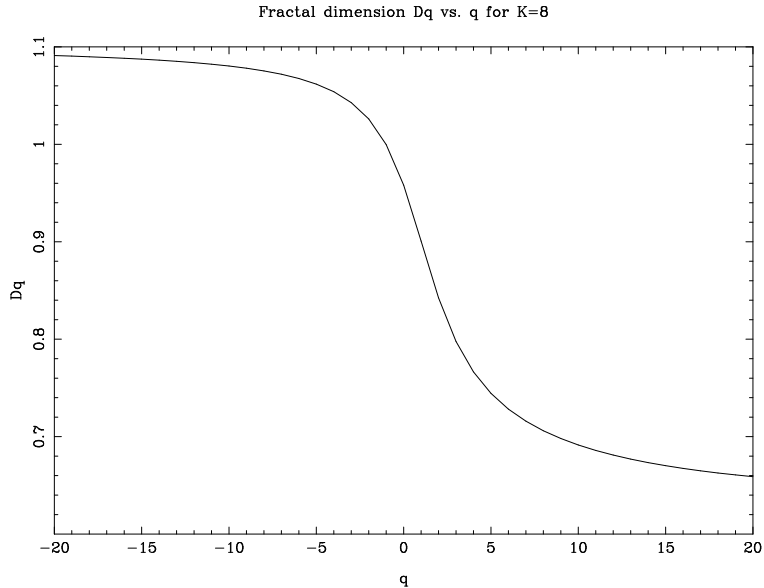}}}
\caption{Multifractal R\'{e}nyi Dimension plot (K=8) for the bacteria {\it C.~jejuni}. 
Note that the value $D_{1}$ is the Shannon entropy of the genome for a symbol size of 8. 
The graph is relatively linear in the region $[-2,1]$ which suggest these values of 
$D_{q}$ can be used in vectors in a Euclidean space.}
\label{Dq_plot}
\end{figure}

The multifractal $D(q)$ plot for {\it Campylobacter jejuni~\cite{CJejuni}} 
is shown in Figure~\ref{Dq_plot}. As with the Yu 
{\it et al.} we found that a segment size of $K=8$ works best in classifying
bacteria. The near linearity of the $D(q)$ plot around $q=0$ suggests that we can assign to each
bacteria a point in ${\mathbf R}^{2}$ or ${\mathbf R}^{3}$ given by $(D_{-1},D_{1})$ or $(D_{-1},
D_{1},D_{-2})$. Yu {\it et al.}~found that phylogenetically close bacteria are close in the 
two spaces.
We use the space $(D_{-1},D_{1},D_{-2})$ in conjunction with the minimal-span tree 
algorithm~\cite{Winter} to generate phylogenetic trees in the following subsection.
\subsection{Phylogenetic trees}
\begin{figure}[htb]
  \centering
  \subfigure[Phylogenetic tree obtained using the multifractal metric]{
    \label{ptree:mf}
    \includegraphics[height=5.5cm]{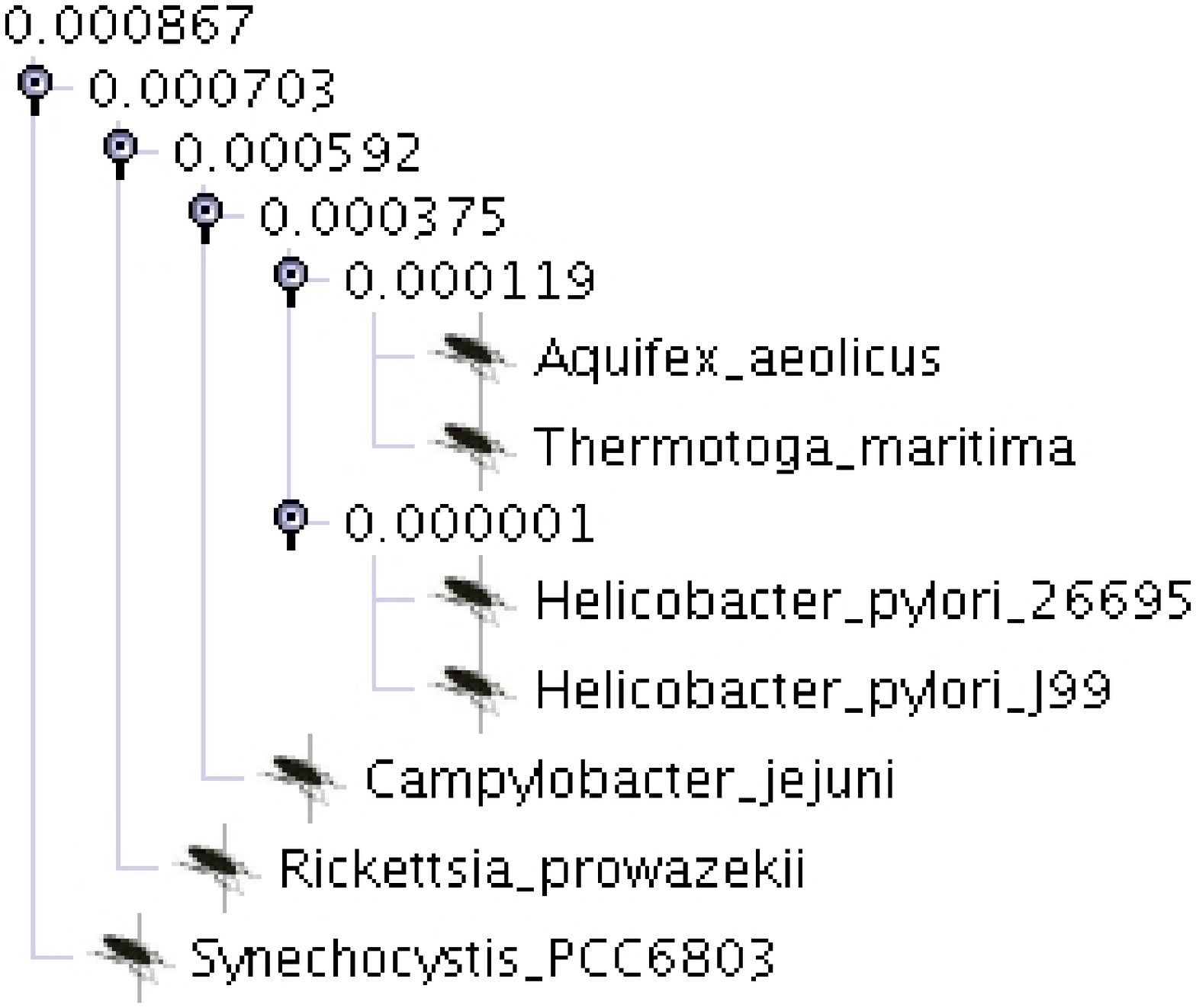}
  }
  \subfigure[Phylogenetic tree obtained using the chi-squared metric]{
    \label{ptree:cs}
    \includegraphics[height=5.5cm]{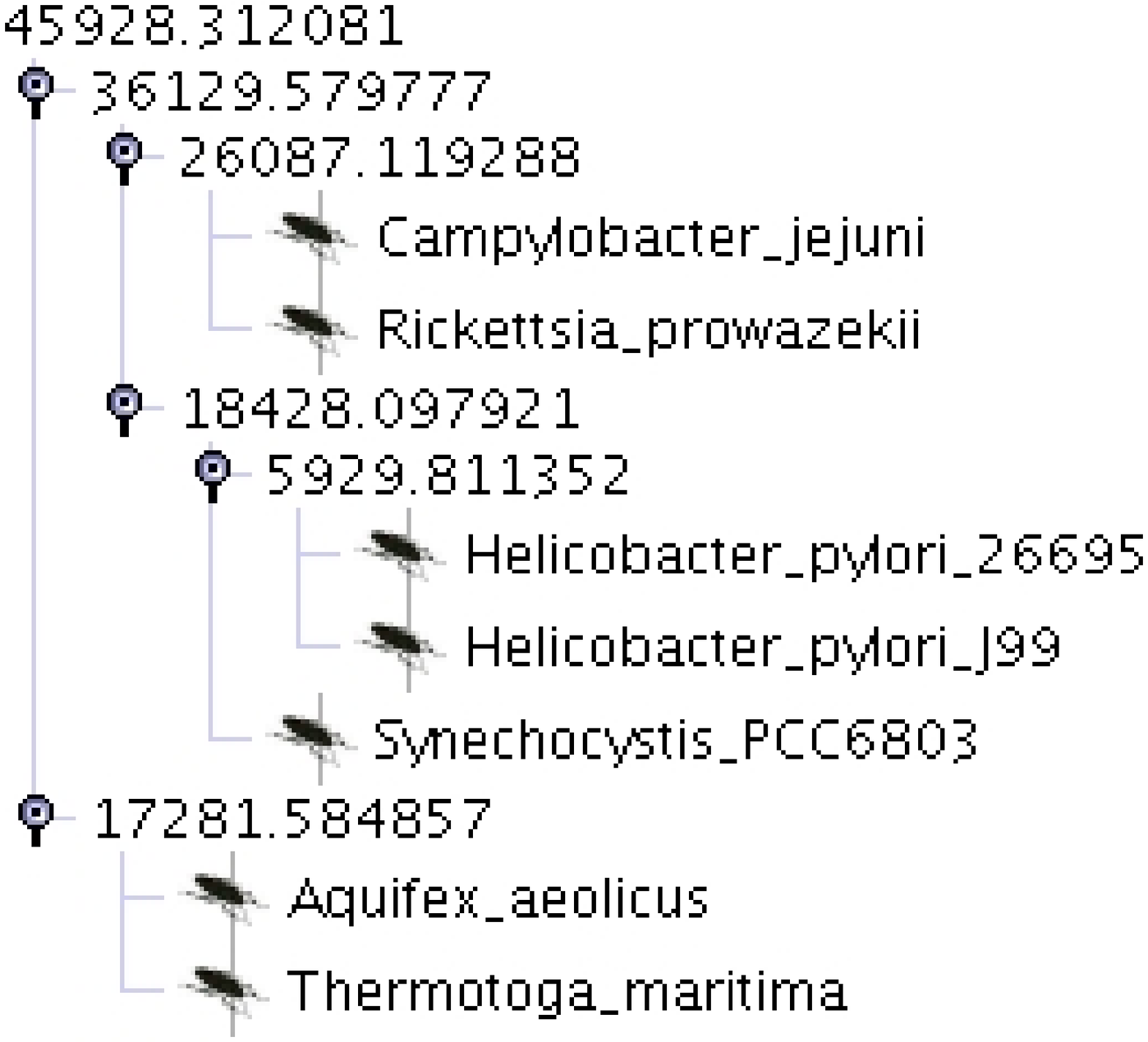}
  }
  \caption{The result of applying the minimal-span tree algorithm to the mulitfractal distance
    metric in Eq.~\ref{dmetric} is shown for several members of the proteo-bacteria family
    in Fig.~\ref{ptree:mf}. Using the chi-squared metric in Eq.~\ref{chi} instead results in
    the tree shown in Fig.~\ref{ptree:cs}.
    The miniature bug icons represent the organisms we see today, the circles represent the
    branches of the tree (where our software thinks the species diverged), and the numbers
    represent the metric scores used to separate the families
    of bacteria at that point.
    Clearly the two {\it H. pylori}~\cite{HPylori} strains group together correctly
    for both metrics.
    A comparison with trees obtained by a detailed analysis of 
    proteins~\cite{protein_trees}, indicates the {\it Thermatoga maritima}~\cite{TMaratima}
    and {\it Aquifex aeolicus}~\cite{AAeolicus} as also closely related, and indeed
    these group together in our trees. 
    Of the two trees, the one using the chi-squared metric appears more correct when
    compared with ones generated from the more usual metrics and tree algorithms
    used in the study of phylogenetic 
    relationships~\cite{protein_trees,gene_content}.}
  \label{ptree}
\end{figure}

For each pair $({\mathbf x},{\mathbf y})$ of genomes, we compute the vectors in Euclidean 
${\mathbf R}^{3}$ space
\begin{equation}
{\mathbf r}_{\mathbf x}=\left(D_{-1}({\mathbf x}),D_{1}({\mathbf x}),D_{-2}({\mathbf x})\right),
\label{rx}
\end{equation}
and
\begin{equation}
{\mathbf r}_{\mathbf y}=\left(D_{-1}({\mathbf y}),D_{1}({\mathbf y}),D_{-2}({\mathbf y})\right).
\label{ry}
\end{equation}
Then we compute the metric 
\begin{equation}
d_{\mathbf x \mathbf y}^{2}=
\norm{{\mathbf r}_{\mathbf y}-{\mathbf r}_{\mathbf x}}^{2}
\label{dmetric}
\end{equation}
and use this in the minimal-span tree algorithm~\cite{Winter} to generate binary phylogenetic 
trees. We have used this approach to generate the phylogenetic tree for members of the 
proteo-bacteria and hyperthermophile families of bacteria as shown in Figure~\ref{ptree:mf}.

Another method we have been exploring in relation to both text and DNA is a quantitative 
chi-squared method which computes a metric with lower scores indicating a closer match.
Similar to the inter-word spacing technique for text, for DNA we compute a scaled
standard deviation of spacing, in this case for codons. For example, the spacing
for the codon {\it gat} in the sequence {\it gat agg gcg gat} is two. Note we
simply break the sequence into groups of three bases, starting at the beginning, to form
codons; while not correct in the sense of the true biology of gene reading we ignore this
problem as we are only interested in large scale properties of the sequence. We compute
the scaled standard deviations as per Eq.~\ref{stddev}.

This gives sets of variances of codon spacings for all the genomes, 
$\{\hat{\sigma}^{2}_{11},\ldots,\hat{\sigma}^{2}_{I1}\},\ldots
,\{\hat{\sigma}^{2}_{1J},\ldots,\hat{\sigma}^{2}_{IJ}\}$, for all possible codons,
labelled $i=1,\ldots,M$ and genomes $j=1,\ldots,J$.
Then we use a formula for $\chi^{2}$ as given in Kullback~\cite{Kullback} for a pair of genomes 
$(k,l)\in\{1,\ldots,J\}\times\{1,\ldots,J\}$.
\begin{equation}
\displaystyle \chi^{2}_{kl} = \frac{1}{N_{k}N_{l}} \sum_{i=1}^{I} 
\frac{\left( N_{l} \hat{\sigma}_{ik}^{2} - N_{k} \hat{\sigma}_{il} \right)^{2}}
{\hat{\sigma}_{ik}^{2} + \hat{\sigma}_{il}^{2}},
\label{chi}
\end{equation} 
\begin{equation}
  \displaystyle N_{k}=\sum_{i=1}^{I}\hat{\sigma}_{ik}^{2},
  \label{N1}
\end{equation}
\begin{equation}
  \displaystyle N_{l}=\sum_{i=1}^{I}\hat{\sigma}_{il}^{2}.
  \label{N2}
\end{equation}
We thus generate a set of $\chi^{2}$ values for each pair of genomes.
As with the multifractal metric, the chi-squared values can be combined
with the minimal-span tree algorithm to produce a phylogenetic tree.
For comparison between the trees generated between the metric, see Figure~\ref{ptree}

\section{CONCLUSIONS}
We have presented a number of interesting methods for analyzing both text and DNA.
The graphs of inter-word spacing standard deviation highlight some interesting results,
we are pursuing further work in extracting relevant features of the graphs. As we have mentioned,
a web search engine is currently under development to give a qualitative analysis of 
different keyword extraction techniques.
The color spectrogram technique shows promise, we are working on 
improving the color contrast between important features and the general background colors. 
\acknowledgments
We gratefully acknowledge funding from The University of Adelaide.
\bibliography{projectbib}   
\bibliographystyle{spiebib}   
\end{document}